\documentclass{article}
\usepackage{epsfig,amssymb,natbib}
\usepackage{graphicx}
\usepackage{fullpage}

\let\ssection=\section
\renewcommand{\section}{\setcounter{equation}{0}\ssection}

\def\e{\mathrm{e}}
\def\i{\mathrm{i}}
\def\d{\mathrm{d}}

\def\beq{\begin{equation}}
\def\eeq{\end{equation}}
\def\Re{\mathrm{Re}\,}
\def\Im{\mathrm{Im}\,}
\def\eps{\epsilon}

\def\hom{\hat{\omega}}
\def\hc{\hat{c}}

\newcommand{\inter}[1]{\quad \textrm{#1} \quad}

\newcommand{\dt}[2]{\frac{\mathrm{d} #1}{\mathrm{d} #2}}
\newcommand{\eqn}[1]{(\ref{eqn:#1})}
\newcommand{\lab}[1]{\label{eqn:#1}}
\newcommand{\av}[1]{\langle #1 \rangle}

\title{Unbalanced instabilities of rapidly rotating stratified shear flows}
\author{J. Vanneste$^1$ and I. Yavneh$^2$}

\date{\normalsize{$^1$ School of Mathematics and Maxwell Institute for Mathematical Sciences, University of Edinburgh, Edinburgh EH9 3JZ, UK} \medskip \\ 
\normalsize{$^2$ Department of Computer Science, Technion, Haifa 32000, Israel}}

\begin{document}

\maketitle

\begin{abstract}
The linear stability of a rotating, stratified, inviscid horizontal plane Couette flow in a channel is studied in the limit of strong rotation and stratification. 
Two dimensionless parameters characterize the flow: the Rossby number $\eps$, defined as the ratio of the shear to the Coriolis frequency and assumed small, and the ratio $s$ of the Coriolis frequency to the buoyancy frequency, assumed to satisfy $s \le 1$. 
An energy argument is used to show that unstable perturbations must have large, $O(\eps^{-1})$ wavenumbers. This motivates the use of a WKB-approach which, in the first instance, provides an approximation for  the dispersion relation of the various waves that can propagate in the flow. These are Kelvin waves, trapped  near the channel walls, and inertia-gravity waves with or without turning points. 

Although, the wave phase speeds are found to be real to all algebraic orders in $\eps$, we establish that the flow is unconditionally unstable.  This is the result of linear resonances between waves 
with oppositely signed  wave momenta. Three modes of instabilities are identified, corresponding to the resonance between (i) a pair of Kelvin waves, (ii) a Kelvin wave and an inertia-gravity wave, and (iii) a pair of inertia-gravity waves. Whilst all three modes of instability are active when the Couette flow is anticyclonic, mode (iii) is the only  possible instability mechanism when the flow is cyclonic. 

We derive asymptotic estimates for the instability growth rates. These are exponentially small in $\eps$, of the form $\Im \omega = a \exp(-\Psi/\eps)$ for some positive constants $a$ and $\Psi$. 
For the Kelvin-wave instabilities (i), we obtain analytic expressions for $a$ and $\Psi$; the maximum growth rate, in particular, corresponds to $\Psi=2$.
For the other types of instabilities, we make the simplifying assumption $s \ll 1$ and find that $\Psi=2.80$ for (ii) and $\Psi=\pi$ for (iii). 
The asymptotic results are confirmed by numerical computations. These reveal, in particular, that the instabilities (iii) have much smaller growth rates in cyclonic flows than in anticyclonic flows, in spite of having both $\Psi=\pi$. 

Our results, which extend those of \cite{kush-et-al} and \cite{yavn-et-al}, highlight the limitations of the so-called balanced models, widely used in geophysical fluid dynamics, which filter out Kelvin and inertia-gravity waves and hence predict the stability of the Couette flow. They are also relevant to the stability  of Taylor--Couette flows and of astrophysical accretion discs.
\end{abstract}

\section{Introduction}

Rapid rotation and strong density stratification characterise the
dynamics of geophysical fluids, the atmosphere and the oceans in
particular. Two dimensionless numbers are used to measure the
importance of these two effects relative to nonlinear advection:
the Rossby number
\[
\eps = \frac{U}{fL},
\]
and the Froude number
\[
F = \frac{U}{N D}.
\]
Here $U$ is a typical horizontal velocity, $f > 0$ is the Coriolis
parameter, $N$ the Brunt--V\"ais\"al\"a frequency, and $L$ and $D$ are
typical horizontal and vertical length scales. With $N>f$, as is
realistically the case, the Rossby number estimates
the maximum ratio between the typical frequency of the (slow) advective motion
(given by $U/L$), and the frequency of inertia-gravity waves (bounded
from below by $f$). Its smallness, explicity
$\eps \ll 1$,
has an important dynamical consequence, namely the weakness of the interaction
between advective motion and inertia-gravity waves. This, together
with the observation that inertia-gravity waves have generally weak
amplitudes in the atmosphere and oceans, has led to development ---
and success --- of the
so-called balanced models, which filter out inertia-gravity waves
completely. These models describe only the slow, large-scale
dynamics, termed balanced because of its
closeness to hydrostatic and geostrophic balance. They can be derived
asymptotically, using power-series expansions in $\eps$, and in principle can achieve an arbitrary algebraic accuracy  $O(\eps^n)$ \citep[e.g.,][]{warn97,warn-et-al}.
%

To understand balanced dynamics and its limitations more fully, it is important to identify and quantify the phenomena that balanced models fail to capture. 
Of particular interest are those unbalanced phenomena which occur in spite of the smallness of $\eps$ and cannot be suppressed by balancing the initial data. 
In the present paper we consider one such mechanism, namely the instability of balanced flows to unbalanced, gravity-wave-like perturbations.
Since this type of instability is absent from
balanced models of arbitrary high accuracy (which
all have qualitively similar stability conditions; see
\citealt{ren-shep}), the growth rates can be expected to be $o(\eps^n)$ for all $n \ge 1$ or, in other words, to be beyond all orders  in $\eps$, and typically exponentially small in $\eps$.  Our results confirm this scaling and show that the instability bands, i.e., the range of unstable wavenumbers, are exponentially narrow. 

We note that unbalanced instabilities like the one examined in this paper are distinct from the mechanism of spontaneous generation of inertia-gravity waves 
sudied in \citet{v-yavn04}. Both mechanisms are exponentially weak, but whilst the exponentially small quantity is the growth rate for instabilities, it is the amplitude of the waves in the case of spontaneous generation. This difference may not be essential, however, if the unbalanced instabilities saturate at a level that decreases to zero with growth rate, as is typical. Another difference is the fact that the instabilities require an initial unbalanced perturbation, whilst spontaneous generation occurs from entirely balanced initial conditions.
We emphasize that both mechanisms provide potential sources of inertia-gravity waves in the atmosphere and oceans. What the exponential smallness indicates in both cases is that the effectiveness of these sources is highly sensitive to the Rossby number.

\begin{figure}
\begin{center}
\epsfig{file=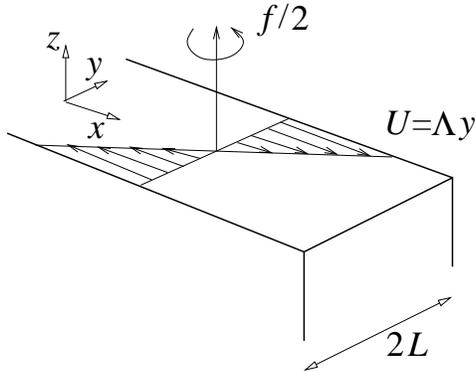,height=5cm}
\caption{Schematic of the Couette flow, with velocity $(\Lambda y,
  0,0)$, of a
  three-dimensional Bousinesq fluid with constant Brunt--V\"ais\"al\"a
frequency $N$. The domain rotates around the (vertical) $z$-axis at
rate $f/2$; it is unbounded in the $x$ and $z$ directions, and bounded in
the $y$ directions by two rigid walls at a distance $2 L$.}
\label{fig:channel}
\end{center}
\end{figure}

The specific flow whose stability we study is a
horizontal Couette flow with velocity $(\Lambda y,0,0)$, modelled
using the Boussinesq approximation with constant $N$, and an
$f$-channel of width $2L$. See Figure
\ref{fig:channel} for an illustration. A natural definition of a (signed)
Rossby number for this flow is the ratio
\[
\eps = \frac{\Lambda}{f}
\]
of (minus) the basic-flow vorticity to the planetary vorticity.
For $\eps > 0$ ($<0$), the shear is anticyclonic (cyclonic).
The other dimensionless
parameter characterising the flow can be taken to be the Prandlt
ratio
\[
s = \frac{f}{N}.
\]
We restrict our attention to $ s \le 1$ and note that in the
atmosphere and oceans $s \ll 1$ generally holds.

Because it is steady, the flow under consideration remains exactly
balanced for all times, unlike generic time-dependent flows.
Furthermore, it is stable in any balanced approximation, however accurate:
this is because the shear is linear, and hence
the potential vorticity constant, whilst balanced instabilities
are inflectional instabilities,which require changes in the sign of
the potential-vorticity gradient. Thus, with this flow, there are none
of the difficulties in separating inertia-gravity waves from balanced
motion that would appear for more complicated flows, and the analysis
reduces to a straighforward linear stability analysis. The smallness
of $\eps$ is of course exploited to derive asymptotic results.

A number of authors have investigated gravity-wave-like instabilities
of shear flows, although mostly in the context of two-dimensional
(shallow-water or compressible-gas) models, in either
parallel or cylindrical geometry \citep{sato81a,sato81b,
  sato82,nara-et-al,knes-kell,ford94a,balm99,drit-v}, and of isentropic models \cite[][and references therein]{papa-prin}. The emphasis
was not, however, on the small $\eps$ limit; indeed, in
shallow water, flows with $\eps \ll 1$ and $F=O(\eps)$ are
linearly stable as Ripa's theorem indicates \citep{ripa83}. In contrast, the
three-dimensional model examined here turns out to be always unstable,
with growing modes whose horizontal and vertical wavenumbers scale
like $\eps^{-1}$. Our analysis has nevertheless many common features
with some of the works cited above, in
particular the use of the WKB approximation. A common theme \citep[in
particular with][]{nara-et-al} is also
the interpretation of the instabilities in terms of (linear)
resonances between
modes with differently signs of the conserved wave energy (or
pseudoenergy) and wave momentum (or pseudomomentum) \citep[see,
e.g.,][and references therein]{craik,ripa90}.

In the presence of lateral boundaries, as is the case here, there are
two types of unbalanced modes: inertia-gravity waves, which are
oscillatory in the cross-stream direction, and Kelvin waves, which
are trapped at each boundary. Instabilities involving the
resonance of Kelvin waves have been studied recently by
\citet{kush-et-al} for the model considered here, and by
\citet{yavn-et-al} and \citet{mole-et-al01} in the annular geometry of the (stratified) Taylor--Couette flow (see also \citet{rudi-et-al} and \citet{dubr-et-al} for astrophysical applications). For simple geometric reasons,
these instabilities occur only for anticyclonic shears ($\Lambda>0$). 
\citet{yavn-et-al} and \citet{mole-et-al01} also identified other modes of instability in anticyclonic shears. These can be associated with the resonance between Kelvin and inertia-gravity waves, and between inertia-gravity waves. The first mechanism is analogous to the mixed-mode instabilities examined by \citet{saka89}, \citet{mcwi-et-al04}, \citet{mole-et-al05} and \citet{plou-etal} in a variety of contexts. As we show, the second mechanism is also active in cyclonic shears ($\Lambda<0$). Thus, we establish that the stratified horizontal Couette flow is unconditionally unstable.

For all the instabilities that we study, the growth rates are exponentially
small in $\eps$ because the resonant waves with differently signed wave
momentum are localised exponentially in different sides of the
channel. We provide both a qualitative description of the
instabilities, based on the mode resonance and conservation laws, and
quantitative results, based on the WKB approximation and numerical
computations.

The remainder of this paper is organized as follows. The linearized equations of motion governing the evolution of perturbations in the Couette flow are introduced in \S\ref{sec:model}. The conservation laws for the wave momentum and wave energy are also introduced there. The latter conservation law is used to show that the horizontal and vertical wavenumbers of growing perturbations must be $O(\eps^{-1})$ or larger. This motivates the WKB approach developed in \S\S\ref{sec:normalmodes}--\ref{sec:inst}. In \S\ref{sec:normalmodes} we formulate the eigenvalue problem for the normal modes of the system, then provide an approximate solution using a WKB expansion (\S\ref{sec:wkb}). To all orders in $\eps$, this leads to purely real eigenfrequencies or, in other words, to waves rather than growing modes. Instabilities with growth rates beyond all orders in $\eps$ are however possible, and we go on to show that they do occur. 
Focusing on the modes susceptible to be involved in instabilities, we give some details of the dispersion relation and structure of Kelvin waves (\S\ref{sec:kelvin}) and inertia-gravity-waves with turning points (\S\ref{sec:igw}). 
We then use arguments based on wave-momentum signature to show that the linear resonance between waves does lead to instabilities for both cyclonic and anticyclonic shears (\S\ref{sec:momsig}). Section \ref{sec:inst} is devoted to the estimation of the instability growth rates. A detailed asymptotic estimate for Kelvin-wave instabilities, extending those of \citet{kush-et-al} and \citet{yavn-et-al}, is derived in \S\ref{subsec:KW-KW}. Rough estimates (focusing on the exponential dependence and ignoring order-one prefactors) are then obtained for the weaker types of instabilities (\S\S\ref{subsec:KW-IGW}--\ref{subsec:IGW-IGW}). These estimates are confirmed by the numerical solutions of the eigenvalue problem presented in \S\ref{sec:numerics}. The paper concludes  with a Discussion in \S\ref{sec:discussion}.

\section{Model} \label{sec:model}

We consider small-amplitude perturbations to the Couette flow described
in the Introduction and in Figure \ref{fig:channel}.
The corresponding linearized equations of motion can be written as
\begin{eqnarray}
D_t u - (f-\Lambda) v &=& - \partial_x p, \lab{lin1} \\
D_t v + f u &=& - \partial_y p, \\
D_t w + \rho &=& - \partial_z p, \\
D_t \rho - N^2 w &=& 0, \\
\partial_x u +\partial_y v + \partial_z w &=&0 \lab{lin5},
\end{eqnarray}
where  $(u,v,w)$ are the
components of the velocity perturbation, $p$ is
pressure perturbation, $\rho$ the buoyancy perturbation, and $D_t = \partial_t +
\Lambda y \partial_x$.
The material conservation
\[
D_t q = 0
\]
of the perturbation potential vorticity
\[
q = (f - \Lambda) \partial_z \rho - N^2 (\partial_x v - \partial_y u)
\]
follows readily. We restrict our attention to perturbations with
vanishing potential vorticity, $q=0$, since this is a characteristic
of unbalanced motion. (See \citet{v-yavn04} for a study of
the generation of inertia-gravity waves from perturbations with
$q\not=0$.) With this restriction, the conservations of the wave energy
(pseudoenergy)
\beq \lab{energy}
\mathcal{E} = \int \! \! \int \! \! \int \left(\frac{|\mathbf{u}|^2}{2} +
\frac{\rho^2}{2 N^2} +
\Lambda y \frac{u \partial_z \rho  -w \partial_x \rho}{N^2} \right) \,
\d x \d y \d z
\eeq
and of the wave momentum (pseudomomentum)
\beq \lab{moment}
\mathcal{M} = \int \! \! \int \! \! \int \frac{ u
\partial_z \rho - w \partial_x \rho}{N^2} \,  \d x \d y \d z
\eeq
are readily derived, as detailed in Appendix \ref{app:a}.

The conservation of  $\mathcal{E}$ constrains the structure
of unstable
perturbations. This is because exponentially growing modes must
have vanishing $\mathcal{E}$ \citep[see, e.g.,][]{ripa90}.
Completing the squares in \eqn{energy}, we rewrite $\mathcal{E}$ as
\begin{eqnarray*}
\mathcal{E} &=& \frac{1}{2} \int \! \! \int \! \! \int \left[
  \left(u+\frac{\Lambda y \partial_z \rho}{N^2}\right)^2
+ v^2 + \left(w - \frac{\Lambda y \partial_x \rho}{N^2} \right)^2
\right. \\
&& \qquad \qquad \qquad \qquad \qquad \left. +
\frac{\rho^2}{N^2} - \frac{\Lambda^2 y^2}{N^4}\left( (\partial_x
  \rho)^2 + (\partial_z \rho)^2 \right)
\right] \, \d x \d y \d z.
\end{eqnarray*}
Clearly, instability can only occur if the perturbation satisfies
$
\Lambda^2 y^2 \left[\partial_x
  \rho)^2 + (\partial_z \rho)^2 \right] > \rho^2 N^2
$
somewhere in the channel. In terms of horizontal and
vertical wavenumbers $k$ and $m$, this gives the condition
\beq \lab{subs1}
\frac{N}{(k^2 + m^2)^{1/2}} < \Lambda L,
\eeq
which can be recognized as a subsonic condition: instability occurs only
for modes whose phase speed is less than the maximum basic-flow
velocity. With $s \le 1$ as assumed, the subsonic condition implies that
$L (k^2 + m^2)^{1/2} \ge \eps^{-1}$, and therefore that modes involved
in instabilies have asymptotically large wavenumbers. One
interpretation of this result states that the Rossby number based on
the wave scale, that is, $\Lambda L (k^2+m^2)^{1/2}/f$, is greater than
unity for unstable modes.

We note that for the shallow-water model with depth $H$, the
subsonic condition analogous to \eqn{subs1} is  $(g H)^{1/2} < \Lambda
L$ and  does not involve the wavenumbers \citep{ripa83}. It is never satisfied for 
sufficiently small $\Lambda$ and thus, for order-one Burger number,
for sufficiently small
$\epsilon$. Thus the shallow-water analogue of our model is linearly
stable in the limit $\eps \to 0$.

\section{Normal modes} \label{sec:normalmodes}

Let us now consider normal-mode solutions of the linearized equations
of motion \eqn{lin1}--\eqn{lin5}. The subsonic condition \eqn{subs1}
suggests that
the wavenumbers $k$ and $m$ should be rescaled by $\eps$.
We therefore
write the dependent variables in the form
\beq \lab{normal}
u(x,y,z,t) = \hat{u}(y/L) \exp \left[ \i \eps^{-1} L^{-1} ( k x + m z
  /s) - f \omega t \right],
\eeq
with similar expressions for $v,w,p$ and $\rho$. Here $k$,
$m$ and $\omega$ are  dimensionless wavenumbers and frequency, with their
dimensional counterparts given by $k/(\eps L)$, $m/(\eps s L)$ and $f
\omega$, respectively. Without loss of generality we
assume that $k>0$. Note that the non-dimensional\-isation then implies
that modes with $\omega > 0$ ($\omega<0$) propagate to the right
(left) in anticyclonic shear and to the left (right) in cyclonic shear.

In terms of the dimensionless $k$ and $m$, the
subsonic condition \eqn{subs1} reads
\beq \lab{subs2}
r = (s^2 k^2+m^2)^{1/2} > 1.
\eeq

Introducing the normal modes \eqn{normal} into
\eqn{lin1}--\eqn{lin5} leads to a system of ordinary differential
equations for $\hat{u}$, $\hat{v}$, $\hat{w}$, $\hat{p}$
and $\hat{\rho}$. These independent variables can be eliminated in
favour of $\hat{p}$, leading in particular to
\beq \lab{uwrp}
\hat{u} = \frac{1}{\eps fL} \frac{\eps (1 - \eps) \hat{p}' +  k \hom
  \hat{p}}{\hom^2-1+\eps}, \quad \hat \rho = - \frac{\i N}{\eps fL}
 \frac{m}{1-s^2 \hom^2}
\hat{p} \quad
\textrm{and} \quad  \hat w = - \frac{1}{\eps NL} \frac{m \hom }{1-s^2
  \hom^2} \hat{p},
\eeq
where prime denotes differentiation with respect to the dimensionless
variable $y/L$ which we henceforth denote simply by $y$. A
second-order differential equation, already obtained by
\citet{kush-et-al}, then follows. It reads
\beq \lab{k1}
\eps^2 \hat{p}'' - \frac{2 \eps^2 k \hom}{1-\hom^2-\eps} \hat{p}' -
\left( k^2 \frac{1-\hom^2+\eps}{1-\hom^2-\eps} + m^2
\frac{1-\hom^2-\eps}{1-s^2 \hom^2}\right) \hat{p} = 0,
\eeq
where
\[
\hom=\omega-k y.
\]
It is supplemented by the boundary conditions $\hat v = 0$, that is, 
\beq \lab{bc}
\eps \hc \hat{p}' + \hat{p} = 0 \quad \textrm{at} \ \ y=\pm 1,
\eeq
where $\hat{c} = c - y = \omega / k - y$.
Note that the singularities of \eqn{k1} for $\hom^2 = 1- \eps$ are removable: in
particular, they are absent from the equation for $\hat{u}$ equivalent
to \eqn{k1} and given in Appendix \ref{app:u}.

\subsection{WKB approximation} \label{sec:wkb}

Together, \eqn{k1} and \eqn{bc} constitute an eigenvalue problem from
which the dispersion relation giving $\omega$ as a function of $k$ and
$m$ can be derived. Taking advantage of the small parameter $\eps$,
this eigenvalue problem can be solved approximately using the WKB method.
To this end, we first expand \eqn{k1} in powers of
$\eps$, with the frequency
\[
\omega = \omega_0 + \eps \omega_1 + \cdots
\]
turning out to be real to all orders.
Taking into account that $\hat{p}'=O(\eps^{-1})$ and
$\hat{p}''=O(\eps^{-2})$, we rewrite \eqn{k1} as
\beq \lab{k2}
\eps^2 \hat p'' - \lambda^2 \hat p -  \frac{2 \eps^2k
  \hom_0}{1-\hom_0^2} \hat p' +
  \eps  h \hat p = O(\eps^2),
\eeq
where
\beq \lab{lambda}
\lambda^2 = k^2 + m^2 \frac{1-\hom_0^2}{1-s^2 \hom_0^2}.
\eeq
and
\[
h =\frac{2k^2}{\hom_0^2-1} + \frac{m^2}{1-s^2 \hom_0^2}+ 2
\omega_1 \frac{m^2 (1-s^2) \hom_0}{(1-s^2 \hom_0^2)^2}.
\]
We introduce solutions of the form
\beq \lab{wkb}
\hat p = (g_{\pm} + \eps g_{1\pm} + \cdots) \exp \left[\pm |\eps|^{-1} \int^y
  \lambda(y') \, \d
  y'\right]
\eeq
into \eqn{k2} and find that $g_\pm$ satisfies
\beq \lab{g'/g}
\frac{g_{\pm}'}{g_\pm} = - \frac{\lambda'}{2 \lambda} + \frac{k
  \hom_0}{1-\hom_0^2} \mp \sigma \frac{h}{2\lambda},
\eeq
where $\sigma=\mathrm{sgn} \, \eps$ equals $+1$ for an anticyclonic shear
and $-1$ for a cyclonic shear. The solution can be written as
\beq \lab{gpm}
g_\pm = A \left(\frac{1-\hom_0^2}{\lambda}\right)^{1/2} \exp \left[ \mp
  \sigma \int^y \frac{h(y')}{2 \lambda(y')} \d y' \right],
\eeq
where $A$ is an arbitrary complex constant. Note that this solution is
single-valued near $\hom_0 = \pm 1$, consistent with the observation
that the singularites of \eqn{k1} for $\hom^2 = 1 - \eps$ are
removable: the multi-valuedness caused by the
square root factor in \eqn{gpm} is cancelled by that of the integral
in the argument of the exponential.

We can classify the solutions \eqn{wkb} according to the sign of $\lambda^2$ in
the channel and distinguish:
\begin{description}
\item[Kelvin waves (KWs)], for which $\lambda^2 > 0$ for
  $-1 \le y \le
  1$. These modes are trapped exponentially near one of the boundary,
  with $O(\eps)$ trapping scale.
\item[Inertia-gravity waves (IGWs)], which satisfy $\lambda^2 <
  0$ in at least part of the channel. There they have an oscillatory
  structure with $O(\eps)$ wavelength.
\end{description}

We now derive approximate dispersion relations for both types of
waves. Together with information on the signature of their wave
momentum discussed in \S\ref{sec:momsig}, these allow the prediction of
instabilities associated with KW-KW, KW-IGW and IGW-IGW resonances.
Asymptotic estimates for the growth rates of these instabilities are
derived in \S\ref{sec:inst}, where they are compared with numerical
results.

\subsection{Kelvin waves} \label{sec:kelvin}

We first consider WKB solutions to \eqn{k1} for which $\lambda^2>0$.
Two independent solutions can be written as
\begin{eqnarray}
p_- &=& \left(\frac{1-\hom_0^2}{\lambda}\right)^{1/2}  \exp \left\{-
    |\eps|^{-1}  \int_{-1}^y \left[\lambda(y')-\eps
    \frac{h(y')}{2 \lambda(y')}\right] \d y' \right\}
\left[1+O(\eps)\right]
\lab{p-} \\
p_+ &=&  \left(\frac{1-\hom_0^2}{\lambda}\right)^{1/2} \exp \left\{ -
    |\eps|^{-1}  \int_y^1 \left[\lambda(y')-\eps
    \frac{h(y')}{2 \lambda(y')}\right] \d y' \right\}
\left[1+O(\eps)\right]. \lab{p}
\end{eqnarray}
The dispersion relation is found from the boundary conditions in the
form
\beq \lab{dispfull}
\left| \begin{array}{cc}
\eps \hc(-1) p'_-(-1)+ p_-(-1) & \eps \hc(-1)  p'_+(-1)+p_+(-1) \\
\eps \hc(1)  p'_-(1)+  p_-(1) & \eps \hc(1) p'_+(1)+ p_+(1)
\end{array}
\right| = 0.
\eeq
Since the off-diagonal terms are exponentially small, the dispersion
relation factorises to all orders into two branches corresponding to
KWs trapped at each boundary. We denote by KW$_{\pm}$ the branch
trapped at $y=\pm 1$, respectively; the corresponding frequency satisfies
\beq \lab{dispex}
\eps \hc(1) p'_+(1)+ p_+(1) = 0 \quad  \mathrm{and} \quad \eps
\hc(-1) p'_-(-1)+ p_-(-1) = 0 .
\eeq

At leading order in $\eps$, these two relations reduce to
\[
\sigma \hc_0(1) \lambda(1)  + 1 = 0 \quad \mathrm{and} \quad
  -\sigma \hc_0(-1) \lambda(-1)  + 1 =0.
\]
with solutions
\beq \lab{om0}
 \hc_0(1) = - \frac{\sigma}{r} \quad \mathrm{and}
 \quad \hc_0(-1) = \frac{\sigma}{r}.
\eeq
(In addition, there are spurious solutions $k \hc_0=\pm
\sigma$.)
 Thus, the KWs localised near $y = \pm 1$ have the
leading-order dispersion relation
\beq \lab{c0}
c_0 =  1 -\frac{\sigma}{r} \quad \mathrm{and}
\quad c_0 = - 1 + \frac{\sigma}{r}.
\eeq
 Higher-order approximations for the KW dispersion relation
can be obtained by pursuing the expansion in powers of $\eps$, each
leading to a purely real correction to \eqn{om0}.



\subsection{Inertia-gravity waves} \label{sec:igw}

In the region where the IGW is oscillatory, two independent WKB solutions \eqn{wkb} can be written as
\beq \lab{pi}
p = \left(\frac{1-\hom_0^2}{\ell}\right)^{1/2} \exp \left\{ \pm \i |\eps|^{-1}
\int^y \left[ \ell(y') + \eps \frac{h(y')}{2 \ell(y')} \, \d y'
\right] \right\},
\eeq
where $\ell > 0$ is defined by
\[
\ell^2 = - \lambda^2.
\]
Depending on the value of $\omega$, IGWs can have
at most two turning points, i.e.\ points where
$\lambda=\ell=0$, in the channel. These are located at
\beq \lab{tpoints}
y_\pm = c_0 \pm \frac{1}{r} (1+\delta^2)^{1/2}, \quad \textrm{where}
\quad \delta=m/k,
\eeq
on either side of the `critical level' $y=c_0$ where $\hom_0=0$.
The mode structure is then oscillatory for $y< y_-$ and $y>y_+$, and
exponential for $y_- < y < y_+$. Here, we concentrate on modes with at
least one turning point since, as argued in \S\ref{sec:momsig} below, the
presence of a turning point is necessary for instability.
These IGWs are localised on one side of the channel and exponentially
small on the opposite boundary.

Let us consider one such IGW that is decaying exponentially with $y$
in $[y_-,y_+]$
and denote the corresponding solution by $p_-$. (Its
counterpart, growing exponentially in $[y_-,y_+]$ and
denoted by $p_+$, is readily deduced using
the symmetry $(y,c) \mapsto (-y,-c)$.) In $[y_-,y_+]$, the solution $p_-$
can be written as
\beq \lab{p--}
p_- \sim A \left(\frac{1-\hom_0^2}{\lambda}\right)^{1/2}  \exp \left\{-
    |\eps|^{-1}  \int_{y_-}^y \left[\lambda(y')-\eps
    \frac{h(y')}{2 \lambda(y')}\right] \d y' \right\}.
\eeq
The boundary condition \eqn{bc} at $y=1$ is satisfied automatically to all orders in $\eps$.
The form \eqn{p--} breaks down in an $\eps^{2/3}$ neighbourhood of $y_-$, where
it is replaced by an Airy function $\mathrm{Ai}$. In $[-1,y_-]$, the
solution is given by a linear combination of the two solutions in
\eqn{pi}. The connection formula, which relates the two arbitrary
constants to $A$ and is
found by matching with the Airy function, gives (cf.\ Bender \&
Orszag, Eq.\ (10.4.16))
\beq \lab{p--2}
p_- \sim 2 A \left(\frac{1-\hom_0^2}{\ell}\right)^{1/2} \sin \left\{ |\eps|^{-1}
\int_y^{y_-} \left[ \ell(y') + \eps \frac{h(y')}{2 \ell(y')} \right] \, \d y'
 + \frac{\pi}{4}\right\}.
\eeq
The dispersion relation is then found by applying the
boundary condition \eqn{dispex}  at $y=-1$, leading to
\[
-\sigma \hc(-1) \ell(-1) \cos S(y_-) + \sin S(y_-) = O(\eps),
\]
where
\[
S(y_-)=|\eps|^{-1}
\int_{-1}^{y_-} \left[ \ell(y') + \eps \frac{h(y')}{2 \ell(y')}
\right] \, \d y' + \frac{\pi}{4}.
\]
Solving for $S(y_-)$, we find
\[
S(y_-) = n \pi + \tan^{-1} [\sigma \hc(-1) \ell(-1)] + O(\eps)
\]
where $n$ is an integer. At leading order this gives
\beq \lab{igw0p}
\int_{-1}^{y_-}  \ell(y) \, \d y = n \pi |\eps|,
\eeq
which determines $c_0$ implicitly. The next order relation
determines $c_1$.

Let us write the dispersion relation \eqn{igw0p} for $c_0$ in a
convenient form.
Define $\mu$ by
\[
c_0= -1 + \frac{(1+\delta^2)^{1/2}}{r}  (\mu+1)
\]
where $\delta=m/k$, so that
\[
y_-=-1 + \frac{(1+\delta^2)^{1/2}}{r}  \mu.
\]
The assumption that this turning point is inside the channel imposes
the restriction $0<\mu<2 r (1+\delta^2)^{-1/2}$.
Introducing the integration variable $Y$, with $y=-1+(1+\delta^2)^{1/2}
Y/r$, reduces \eqn{igw0p} to the expression
\beq \lab{mu}
\int_0^\mu \left[ \frac{(Y-\mu-1)^2-1}{1-\nu^2 (Y-\mu-1)^2} \right]^{1/2} \d
Y = \frac{(s^2+\delta^2)^{1/2}}{1+\delta^2} n \pi |\eps|,
\eeq
where
\[
\nu^2 = \frac{s^2(1+\delta^2)}{s^2+\delta^2}.
\]
This defines implicitly a function $\mu(\delta,s,n|\eps|)$ with values
in $[0,\nu^{-1}-1]$, from which $c_0$ is deduced. Taking both the
solution $p_-$ and its symmetric $p_+$ into account, we find the two
branches \beq \lab{igw0} c_0 = \pm 1 \mp \frac{(1+\delta^2)^{1/2}}{r}
[\mu(\delta,s,n |\eps|)+1], \eeq corresponding to modes exponentially
small near $y=\mp 1$ and denoted by IGW$_\pm$, respectively. Again,
higher-order approximations to the phase velocity can in principle be
computed, leading to real corrections to $c_0$ in powers of $\eps$.
Note that, at leading order in $\eps$, the dispersion relation is the
same for both signs of $\eps$, that is, for both cyclonic and
anticyclonic flows. An asymmetry only appears at higher order.

For $n=O(1)$, $\mu \to 0$ as $\eps \to 0$, and the leading-order
dispersion relation reduces to 
\beq \lab{igws} 
c_0 \sim \pm 1 \mp
\frac{(1+\delta^2)^{1/2}}{r}, 
\eeq
corresponding to $y_\pm \to \pm
1$. The small-$\mu$ behaviour of the left-hand side of \eqn{mu}
then suggests that the successive branches $n=1,\, 2, \cdots$ are
$O(\eps^{2/3})$ apart. 

The asymptotic results \eqn{c0} and \eqn{igw0} provide a first
approximation to the dispersion relation of KWs and IGWs. We have
extended this by solving the eigenvalue problem \eqn{k1}--\eqn{bc} (or rather the equivalent formulation \eqn{u}--\eqn{bcu} in terms
of $\hat{u}$) numerically. 
Our numerical solver is the same as the one used in
\citet{yavn-et-al}, employing a second-order finite-volume
discretization of \eqn{u}--\eqn{bcu}. For given physical
parameters and wavenumbers, $m$ and $k$, we search for
eigenfrequencies for which the matrix representing the discretized
system is singular. The codes are
implemented in MATLAB, with the search performed using the {\it fminsearch}
function that employs the so-called Simplex algorithm.


\subsection{Dispersion relation} \label{subsec:disprel}

\begin{figure}
\begin{center}
\includegraphics[width = .7\textwidth]{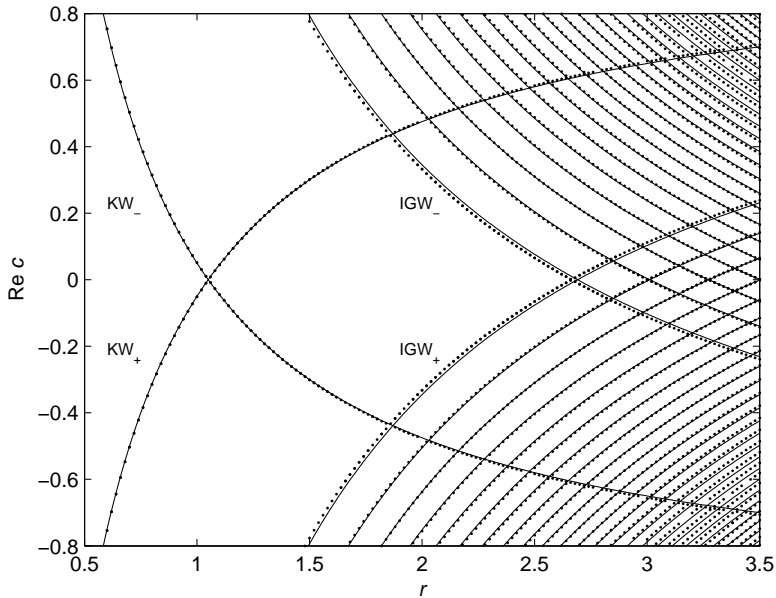}
\caption{Dispersion relation for an anticyclonic flow, with $\epsilon = 0.1$, $\delta=2$ and $s=0.1$.
The two Kelvin waves (labelled KW) are shown along with many inertia-gravity
waves (labelled IGW). The asymptotic estimates for $|\eps|\ll 1$ (solid curves) are compared with numerical solutions (dotted curves).} \label{fig:dispa}
\end{center}
\end{figure}

\begin{figure}
\begin{center}
\includegraphics[width = .7\textwidth]{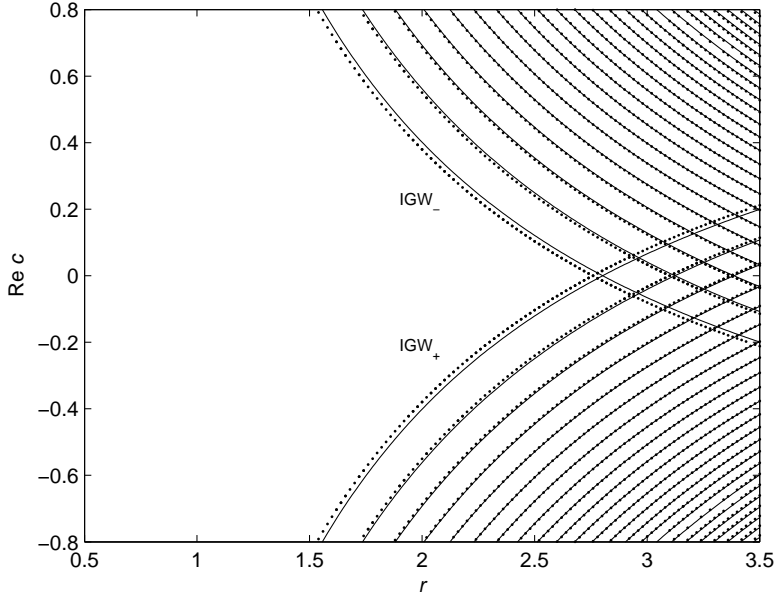}
\caption{Dispersion relation for a cyclonic flow, with $\epsilon=-0.1$. The other parameters and the notation are the same as in Figure \ref{fig:dispa}.} \label{fig:dispc}
\end{center}
\end{figure}

Figures \ref{fig:dispa} and \ref{fig:dispc} show the dispersion
relation for anticyclonic and cyclonic flows, respectively. The parameters have been chosen as $\eps=0.1$, $\delta=2$ and $s=0.1$, but the qualitative features remain the same for a wide range of values. The numerical results (dotted curves) are compared with the asymptotic estimates (solid curves) to confirm the validity of the latter. For KWs, we have used an $O(\eps)$-accurate estimate which improves on \eqn{c0} by adding the term $\eps c_1 = \mp \eps \sigma (2 r)$ derived in Appendix \ref{app:kwinst}. For IGWs, we have used the estimate \eqn{igw0}, corrected in the anticyclonic case by subtracting $\eps/2$ from the square bracket. This correction, which can be viewed as an experimentally determined $O(\eps)$ term in the expansion of $c$, is made for the clarity of the plot: without it, the $O(\eps)$ error in the dispersion relation is not significantly smaller than the $O(\eps^{2/3})$ distance between branches, and it is difficult to relate each asymptotic curve to its numerical counterpart. No corrections were necessary for the cyclonic shear, suggesting the dispersion relation \eqn{igw0} is already $O(\eps)$-accurate in this case. To confirm this would require to continue the asymptotic developments to the next order in $\eps$; this is a daunting task which we have not attempted for IGWs.

Figures \ref{fig:dispa} and \ref{fig:dispc} demonstrate the multiple intersections between the branches
IGW$_{\pm}$ of the dispersion relation. In the anticyclonic case,
there are additional intersections  between the branches
KW$_{\pm}$, and between KW$_\pm$ and IGW$_{\mp}$. (The KW do not appear in Figure \ref{fig:dispc} for the cyclonic case because they have $|c|>1$.)
The
intersections, associated with the linear resonance between modes,
are generically spurious: they result from the finite resolution
of the plot for the numerical results, and from the limited
accuracy for the asymptotic ones. There are in fact two possible
behaviours: (i) mode conversion, when the phase velocities
remaining real and the two curves, rather than intersecting,
locally form the two branches of a hyperbola, or (ii) instability,
when the phase velocities on the two branches become complex
conjugate with non-zero imaginary parts. The two situations are
distinguished by the signs of quadratic invariants, such as the
wave momentum, along the colliding branches: (i) mode conversion
occurs when both signs are the same, and (ii) instability occurs
when the signs differ \citep[e.g.][]{cair79}. We now show that the
latter situation is the relevant one in our problem by examining
the sign of the wave momentum for KWs and IGWs in the WKB
approximation.

\subsection{Wave-momentum signature} \label{sec:momsig}

IGWs and KWs have different leading-order
approximations to
their wave momentum. To see this, we introduce
\eqn{normal} and \eqn{uwrp} into
\eqn{moment} and assume that $\omega$ is real. This gives
\begin{eqnarray}
\mathcal{M} &=& \frac{2N^2 m^2}{f^2 L^2 \eps^3} \int \left[
  \frac{\eps(1-\eps)}{2(\hom^2-1+\eps)(1-s^2 \hom^2)} \dt{|\hat p|^2}{y} +
  \frac{k \hom (1-s^2+s^2 \eps)}{(\hom^2-1+\eps)(1-s^2 \hom^2)^2}
  |\hat p|^2 \right] \, \d
y  \nonumber  \\
&=& \frac{2N^2 m^2}{f^2 L^2 \eps^3} \hat\mathcal{M},\lab{hm}
\end{eqnarray}
where the last line defines the dimensionless wave momentum
$\hat{\mathcal{M}}$ which we will use henceforth.
For IGWs, the first term is negligible: indeed, in
the regions where $p$ oscillates rapidly, $\d |\hat p|^2 / \d y = O(1)$,
while in the possible regions where $p$ decays exponentially, $\d
|\hat p|^2 / \d y = O(\eps^{-1})$ only for a range of $y$ of size $\eps$;
both types of regions thus contribute at $O(\eps)$ to $\hat{\mathcal{M}}$.
This leads to the leading-order approximation
\beq \lab{migw}
 \hat \mathcal{M} \sim \int
  \frac{k \hom (1-s^2) }{(\hom^2-1)(1-s^2 \hom^2)^2} |\hat p|^2  \, \d
y \quad \textrm{for IGWs}.
\eeq
Given that the denominator $(\hom^2-1)$ cancels with the same factor
in $|\hat p|^2$ (see \eqn{wkb}--\eqn{gpm}), it is clear that instability involving IGWs
implies that $\hom$ changes sign. It follows that there is at least one
turning point in the channel, as announced, since the absence of turning points ($\ell^2>0$) implies that $|c|>1$.
Assuming there are turning points, the sign of $\hat
\mathcal{M}$ for the two types of IGWs considered in \S\ref{sec:igw} is then
\[
\hat \mathcal{M} < 0 \ \ \textrm{for IGW}_+ \inter{and}
\hat \mathcal{M} > 0 \ \ \textrm{for IGW}_-.
\]


For KWs, the two terms in \eqn{hm} have a similar,
$O(\eps)$, order of magnitude. Using \eqn{wkb}, we find that
\[
\hat{\mathcal{M}}_\pm \sim
\frac{|\eps|}{2[\hom^2(\pm1)-1][1-s^2\hom^2(\pm1)]} \left[\pm \sigma +
  \frac{k(1-s^2)}{\lambda(\pm1) [1-s^2\hom^2(\pm1)]}
   \right] |\hat{p}(\pm1)|^2.
\]
Using the dispersion relation for Kelvin waves, $\hom(\pm 1) = \mp
\sigma k /r + O(\eps)$ in our non-dimensional\-isation, and its consequence
$\lambda(\pm 1) = r$ (see \eqn{om0} and \eqn{lampm1}), this
reduces to
\[
\hat{\mathcal{M}}_\pm \sim \mp \frac{\eps r^4}{2 m^4}
|\hat{p}(\pm1)|^2 \quad \textrm{for KWs},
\]
leading to the following signs:
\[
\hat \mathcal{M} \lessgtr 0 \ \ \textrm{for KW}_+ \inter{and}
\hat \mathcal{M} \gtrless 0 \ \ \textrm{for KW}_- \quad \textrm{when}
\ \ \eps \gtrless 0,
\]
differing in the anticyclonic and cyclonic cases.


With the wave-momentum signatures just obtained, it is clear from Figures \ref{fig:dispa} and \ref{fig:dispc} that the numerous intersections of branches correspond to waves with oppositely signed
$\hat{\mathcal{M}}$. This establishes the existence of many modes of
instability, both for anticyclonic and cyclonic shears. The main
difference between the differently signed shears is that
instabilities involving KWs only are possible only for
anticyclonic shear.

All the instabilities are associated with the interactions of modes
exponentially localised on different sides of the channel. Therefore
their interaction is exponentially weak and, as a consequence,
the growth rates of the instabilities and range of unstable
wavenumbers are
exponentially small in $\eps$, as anticipated in the Introduction. As
the asymptotic calculations of the next section show,
such small growth rates are somewhat delicate to capture
analytically. However, the interpretation in terms of
interactions of waves with oppositely signed $\hat{\mathcal{M}}$ makes
it possible to predict instability robustly, without
detailed calculations.


\section{Instabilities} \label{sec:inst}

\subsection{KW-KW instabilities} \label{subsec:KW-KW}

We start our study of the weak instabilities associated with mode
interactions by deriving an estimate for the growth rate of the
instability that arises through the resonance of KWs in anticyclonic
shear. This instability has been examined in some detail by
\citet{kush-et-al}  and by \citet{yavn-et-al}. Because it is the strongest instability, with physical relevance in Taylor--Couette and accretion discs \citep[see][]{dubr-et-al}, we present here a complete asymptotic derivation of the  growth rate. For the KW-IGW and IGW-IGW instabilities considered in \S\S\ref{subsec:KW-IGW}--\ref{subsec:IGW-IGW}, we limit the derivation to the exponential behaviour of the growth rate as $\eps \to 0$. The method we now describe could however be applied to these instabilities as well, should a more accurate estimate be needed. 

To obtain the growth of the instability, we need to reconsider the
dispersion relation \eqn{dispfull} in the vicinity of the resonance
point, taking into account exponentially small terms.
Let $c_\star$ and $r_\star$ be the values of $r$ and $c$ at resonance.
By symmetry, $c_\star = 0$.
According to \eqn{c0} (with $\sigma=1$ corresponding to the
anticyclonic shear),
\[
r = r_\star = 1 + O(\eps).
\]
Thus, resonance occurs on an ellipse with semi-axes $1/s$ and $1$ in the
$(k,m)$-plane, and the instability region is an
exponentially small annulus around this ellipse. It is best
parameterized using the polar coordinates $(r,\theta)$, with
\[
s k = r \cos \theta \quad \textrm{and} \quad m = r \sin \theta.
\]
Now, take
\[
c = C \quad \textrm{and} \quad r = r_\star + R,
\]
where $C$ and $R$ are exponentially small.
This can be introduced
into the dispersion relation \eqn{dispfull}; using the fact that
$(c=0,r=r_\star)$ satisfy \eqn{dispex}, a Taylor expansion
leaves only terms that are exponentially small. In the coefficients of
$C$ and $R$ in these terms, we can approximate $r_\star$ by its
leading-order estimate $1$.
Noting that, in this approximation,
\[
\lambda(\pm 1) \approx \lambda_\star(\pm 1) + \frac{(s^2 - \cos^2
  \theta) R \pm \cos^2 \theta (1-s^2) C}{s^2 \sin^2 \theta},
\]
we find the dispersion relation in the form
\beq \lab{diskwkw1}
\left(\frac{s^2-\cos^2 \theta}{s^2\sin^2\theta}\right)^2 \left(R^2 -
  C^2\right) = 4 \e^{-2 \Psi/\eps},
\eeq
where
\[
\Psi = \int_{-1}^1 \left[\lambda_\star(y)-\eps
  \frac{h_\star(y)}{2\lambda_\star(y)}\right] \, \d y,
\]
and the subscript $\star$ indicates evaluation at the resonance point.
A consistent approximation of $\Psi$ requires to include the $O(\eps)$
contribution to $\lambda_\star$ in the first term of the integrand.
To this end, we compute the KW dispersion relation to $O(\eps)$ in Appendix
\ref{app:kwinst} and find that $r_\star=1 + \eps /2 + O(\eps^2)$. This leads to
\[
\Psi = \Psi_0 + \eps \Psi_1,
\]
where
\beq \lab{psi0}
\Psi_0=  \int_{-1}^1 \lambda_0(y) \, \d y
\eeq
with
\[
\lambda_0 =  \left[\frac{\cos^2 \theta (1-y^2)+s^2 \sin^2
    \theta}{s^2(1-\cos^2 \theta y^2)} \right]^{1/2},
\]
and
\beq \lab{psi1}
\Psi_1= \frac{1}{2} \int_{-1}^1 \left[\frac{\lambda_0(y)}{1-\cos^2
    \theta y^2}- \frac{\cos^2 \theta y^2}{\lambda_0(y) s^2 (1-\cos^2
    \theta y^2)}
-\frac{h_0(y)}{\lambda_0(y)} \right] \, \d y,
\eeq
with
\[
h_0(y)= \frac{2 \cos^2 \theta}{\cos^2 \theta y^2 - s^2} + \frac{\sin^2
  \theta}{1-\cos^2 \theta y^2}.
\]
The second integral has to be interpreted as a Cauchy principal value at the
singularities $y = \pm s/\cos \theta$ of $h_0(y)$ when these are in $[-1,1]$.
With this result, the dispersion relation \eqn{diskwkw1}  can be rewritten as 
\beq \lab{kwgr}
C = \left[R^2 - \alpha^2 \e^{-2
      \Psi_0/\eps}\right]^{1/2},
\eeq
where
\[
\alpha = \frac{2 s^2 \sin^2 \theta \, \e^{-
      \Psi_1}}{s^2-\cos^2 \theta}.
\]

Formula \eqn{kwgr} is the first main result of this paper.  It provides the leading-order asymptotics for the growth rate of the KW-KW instability (after multiplication by $k$) as $\eps \to 0$ and for arbitrary $s \le 1$). It also makes evident  the exponential smallness of the growth rate and of the instability-band width. 
Its validity is confirmed in \S\ref{sec:numerics} where it is compared with numerical results. 

The minimum of $\Psi_0$, and hence the maximum growth rate, is attained for $\theta = \pi/2$, for which $\Psi_0 \sim 2$. Thus, at the crude level of exponential dependence on $\eps$, we obtain the estimate
\beq \lab{kwkwro}
\log \Im \omega \sim -\frac{2}{\eps}, \quad \textrm{as} \ \ \eps \to 0,
\eeq
for the largest growth rate $\Im \omega=k \Im C$. Note that because $\theta = \pi/2$ implies that $k=0$ and hence $\omega=0$, the maximum growth rate is in fact achieved for $\theta$ slightly less than $\pi/2$; this does not affect the exponential dependence in \eqn{kwkwro}, however (see below). 

Estimates more precise than \eqn{kwkwro} can of course be inferred from \eqn{kwgr}. Focusing on the limit $\theta \to \pi/2$, we note that $C$ depends on the relationship between $s$ and $\theta$. A distinguished limit is found for $s = O(\cos \theta) \ll 1$. This corresponds to the regime with $s \ll 1$ and $\delta=k/m=O(1)$, which we term the quasi-geostrophic regime, since it  
corresponds to the quasi-geostrophic scaling implying, in particular, the hydrostatic approximation ($k/m$ can be recognized as the square root of the Burger number based on the wave scale). Taking the limit $\theta \to \pi/2$
of \eqn{psi0}--\eqn{kwgr} with $k = s/\cos \theta$ fixed then yields
\[
\Psi_0 \sim 1 + \frac{1+k^2}{k} \tan^{-1} k \inter{and} \alpha \sim 2
  \frac{|1-k|^{k-1}}{|1+k|^{k+1}}.
\]
The maximum of the imaginary part of the phase speed is then
obtained for $k \to 0$ and given by $\Im c \sim 2 \exp(-2/\eps)$, consistent with \citet{yavn-et-al}'s equation (35).  The maximum of the growth rate $\Im \omega$ is easily seen to be attained for $k=O(\eps^{1/2})$ and to be a factor $\eps^{1/2}$ smaller than the maximum of $\Im c$. In dimensional terms, this means that the horizontal and vertical scales are both large, but have different  orders of magnitudes, scaling like $\eps^{-1/2}$ and $\eps^{-1}$, respectively.



\subsection{KW-IGW instabilities} \label{subsec:KW-IGW}

The KW-IGW instabilities occur for anticyclonic flows through the
resonance of an IGW,  which has one
turning point and is localised on one side of the channel, with a KW
localised
on the other side. To estimate their growth rates, we can consider a
solution consisting of a linear combination of the IGW$_-$ given by
\eqn{p--}--\eqn{p--2} which is oscillatory near $y=-1$, and the KW$_+$
given by \eqn{p}. (The other combination, of IWG$_+$ with KW$_-$, has the same growth rate, by symmetry.)
A calculation similar to that carried out for KW-KW
instabilities could in principle be performed to obtain the
leading-order behaviour of the growth rate. However, this requires the
derivation of the IGW dispersion relation accurate to $O(\eps)$ involving an inordinate amount of calculation. We shall therefore
limit ourselves to the determination of the exponential behaviour of
$\Im \omega$ (that is, to the determination of the constant $\Psi_0$ such
that $\log \Im \omega \sim  -\Psi_0/\eps$ as $\eps \to 0$)
in the instability regions, and ignore the order-one prefactor in the expression of $\Im \omega$.
As in the case of KW-KW
instabilities, $\Psi_0$ is determined simply from the
amplitude of the colliding modes at the boundary where they are
exponentially small, given explicitly by
$\exp(-\Psi_0/\eps)$.
Note that $\Psi_0$ controls not only the exponential smallness of the growth
rate but also that of the width of the instability bands.

For simplicity
we restrict our analysis to the quasi-geostrophic scaling $s\ll
1$, $\delta=O(1)$.
For $s \ll 1$ and $\sigma=1$, the phase speeds of colliding KW$_+$ and IGW$_-$
branches given in \eqn{c0} and \eqn{igw0} reduce at leading order to
\[
c = 1 - \frac{1}{m} \inter{and} c = -1 + \left(\frac{1}{k^2} +
  \frac{1}{m^2} \right)^{1/2},
\]
respectively. 
The corresponding resonance condition
\[
\frac{1}{m} +  \left(\frac{1}{k^2} +
  \frac{1}{m^2} \right)^{1/2} = 2,
\]
that is,
\beq \lab{kwigwres}
k = \frac{1}{2}\left(\frac{m}{m-1} \right)^{1/2}, \quad \textrm{with} \ \ m>1
\eeq
defines a curve in the $(k,m)$ plane in the vicinity of which
instabilities are concentrated. For KW-IGW instabilities, since there is a
single turning point $y_-$ in the channel, $\Psi_0$ is given as 
\beq \lab{kwigwpsi}
\Psi_0 =  \int^1_{y_-} \lambda(y) \, \d y.
\eeq
The integrand $\lambda(y)$, given in \eqn{lambda}, can be approximated
by
\beq \lab{kwigwlam}
\lambda(y) = km \left[ (y_+ - y) (y-y_-) \right]^{1/2},
\eeq
with $y_\pm$ reducing to
\beq \lab{kwigwtp}
y_\pm = c \pm \left(\frac{1}{k^2} +
  \frac{1}{m^2} \right)^{1/2} = \left\{ \begin{array}{l} 3 -2/m \\ -1
  \end{array} \right. .
\eeq
Introducing \eqn{kwigwres} and \eqn{kwigwlam}--\eqn{kwigwtp} into
\eqn{kwigwpsi} gives the expression
\[
\Psi_0 = \frac{1}{8[m(m-1)]^{1/2}} \left[(2m-1)^2\left( \pi + 2 \sin^{-1}
  \frac{1}{2m-1}\right) +4(m(m-1))^{1/2} \right].
\]
The maximum growth rate of the KW-IGW instability is given by the
minimum value of $\Psi_0$, found to be
\beq \lab{kwigwkm}
\Psi_0 = 2.80\cdots \quad \textrm{for} \ \ k = 1.04\cdots \ \ \textrm{and} \ \
m=1.30\cdots.
\eeq
Thus we obtain the asymptotics
\beq \lab{kwigwro}
\log \Im \omega \sim -\frac{2.80}{\eps}, \quad \textrm{as} \ \ \eps \to 0,
\eeq
for the growth rate of KW-IGW instabilities.
Comparison with \eqn{kwkwro} then indicates that these are considerably weaker than the KW-KW instabilities.

\subsection{IGW-IGW instabilities} \label{subsec:IGW-IGW}

We now consider the instabilities that result from the resonance
between IGWs. These are particularly important for cyclonic flows
since they provide the only mode of instability in this case.
In fact, as can be expected from the leading-order dispersion relation
\eqn{igw0}, the dominant behaviour of these instabilities is
unaffected by rotation, so that the exponential dependence on $1/\eps$
is identical for anticyclonic and cyclonic shears. What differs
between the two cases, however, is the order-one prefactor which we do not estimate analytically.

IGW-IGW instabilities occur when a solution $p_-$ of the form
\eqn{p--}--\eqn{p--2} is resonant with its counterpart $p_+$.
The modes have then two turning points $y_\pm$ in the channel, leading
to the necessary condition $r \ge (1+\delta^2)^{1/2}$ for the
instability. We now estimate the factor $\Psi_0$ controlling the
exponential smallness of the instability growth rates. As in the
previous section, we restrict our attention to the quasi-gesotrophic
scaling $s \ll 1$ and $\delta=O(1)$. We furthermore consider only the
strongest IGW, associated with the (symmetric) resonance of the
gravest ($n=1$) IGW modes, and for which $c=0$ to all orders in $\eps$. The
resonance condition is therefore
\[
\frac{1}{k^2} + \frac{1}{m^2} =1.
\]
Since for $n=O(1)$, the two turning points are $y_\pm = \pm 1$ at
leading order in $\eps$, $\Psi_0$ is computed as
\[
\Psi_0 = k m \int_{y_-}^{y_+} [(y_+ - y)(y-y_-)]^{1/2} \, \d y =
\frac{\pi km}{2}.
\]
The minimum value is therefore
\beq \lab{igwigwkm}
\Psi_0 = \pi \quad \textrm{for} \ \ k=m=\sqrt{2},
\eeq
and the exponential scaling of the growth rate given by
\beq \lab{igwigwro}
\log \Im \omega \sim - \frac{\pi}{\eps}, \quad \textrm{as} \ \ \eps \to 0,
\eeq
for both anticyclonic and cyclonic flows.
This is exponentially smaller than the growth rate for either the KW-KW or the KW-IGW instabilities \eqn{kwkwro} or \eqn{kwigwro}.

\subsection{Numerical computation of growth rates}
\label{sec:numerics}

We now present comparisons of the growth rate, or rather $\Im c$, computed numerically with the asymptotic results of \S\S\ref{subsec:KW-KW}--\ref{subsec:IGW-IGW}. The numerical method employed is that described in \S\ref{subsec:disprel} where $\Re c$ was considered.
For the small values of $\eps$ examined here, $\Im c$ is very small and the bands of unstable wavenumbers are very narrow,  so that very fine resolution in $y$ is needed to capture $\Im c$ accurately. 
In order to ensure high accuracy, we successively double the grid resolution until results are unchanged to at least four significant digits. This required grids of sizes
ranging from about 250 mesh points for strong or moderate instabilities, to as
many as 16\,000 mesh points for very weak instabilities. This may be improved
upon by using nonuniform grids with high resolution only in regions where the
solution changes fast.
The search for the bands of instabilities in $(k,m)$ is quite delicate, but made possible by the excelllent approximations afforded by the asymptotic results. 

\begin{figure}
\begin{center}
\includegraphics[width = .7\textwidth]{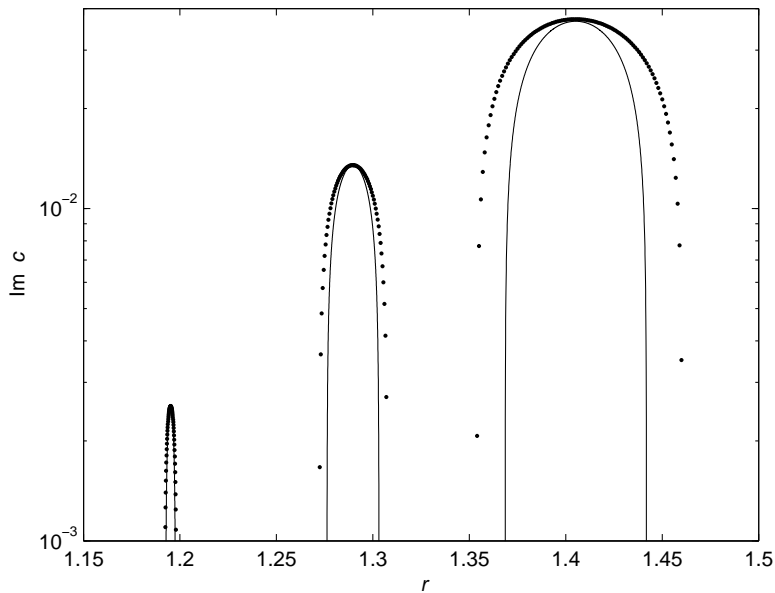}
\caption{Imaginary part of the phase speed ${\Im} c$ as a function of $r$ (in linear--logarithmic scale) for the KW-KW instability in anticyclonic flows with $s=0.1$.  Numerical (dots) and asymptotic (solid curves) results are compared for $\epsilon = 0.3, 0.4, 0.5$ (with increasing $\Im c$), and $\theta = \pi/2$ (i.e.\ $k=0$ and $m=r$).}
\label{fig:profiles}
\end{center}
\end{figure}

\begin{figure}
\begin{center}
\epsfig{file=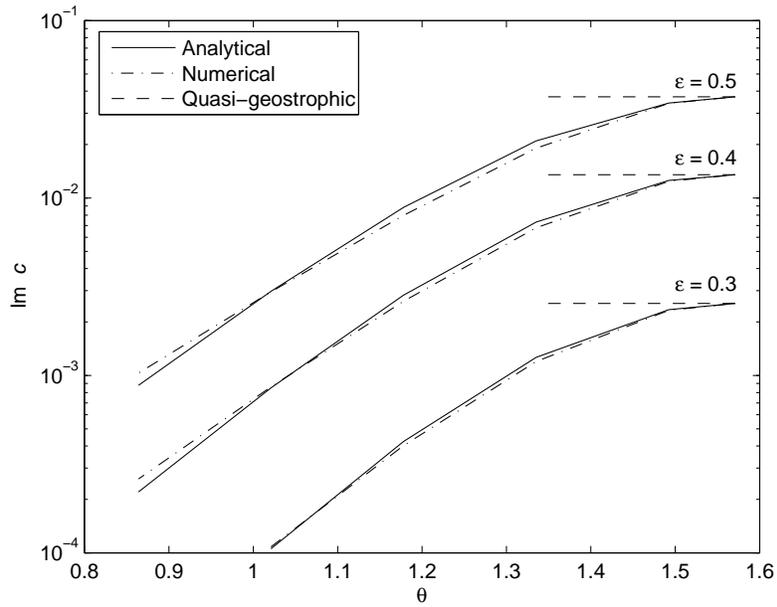,width = .7\textwidth}
\caption{Maximum of ${\Im} c$ as a function of $\theta$ (in linear--logarithmic scale) for the KW-KW instability of anticyclonic flows with $s=0.1$.  Numerical (dash-dotted curves) and asymptotic (solid curves) results are compared for $\epsilon = 0.3, 0.4, 0.5$. The limits of $\Im c$ as $\theta \to 0$, corresponding to the quasi-geostrophic scaling $s \ll 1$ and $\delta=O(1)$ are also indicated.} \label{fig:kwkwtheta}
\end{center}
\end{figure}

We start by considering the KW-KW instability of anticyclonic flows.
Figure \ref{fig:profiles} shows ${\Im} c$ as a function of $r$
for $\theta = \pi/2$ and $\epsilon = 0.3, 0.4, 0.5$ in the
instability bands. The dots represent numerically computed values;
the solid line are computed analytically using (\ref{eqn:kwgr}).
Note that we only know $r_\star$ to algebraic accuracy, while the
bands are exponentially narrow. Hence, we use the numerical
results for determining $r_\star$---the value of $r$ for which
${\Im} c$ is maximized. The narrowing of the instability band
is clearly exhibited in the figure, and the small-$\epsilon$
analytical approximation quickly converges to the numerical
results as $\epsilon$ becomes small. The dependence of $\Im c$ on $\theta$ is illustrated by Figure \ref{fig:kwkwtheta} which compares numerical and asymptotic estimates for the maximum value of $\Im c$ as a function of $\theta$ for $s=0.1$ and $\eps=0.3, \, 0.4$ and $0.5$. The value of $\Im c$ in the quasi-geostrophic scaling $s \ll 1$, $\delta=O(1)$, that is, the limit $\theta \to \pi/2$, is also indicated. The Figure confirms the accuracy of the asymptotic estimate and shows the rapid decrease of $\Im c$ as $\theta$ decrases from $\pi/2$. 

\begin{figure}
\begin{center}
\includegraphics[width = .7\textwidth]{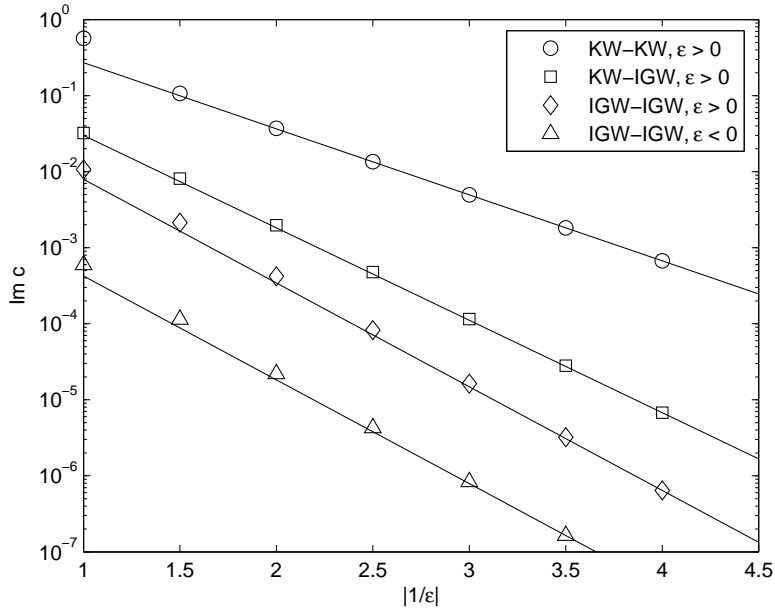}
\caption{Maximum of $\Im c$ as a function of $1/|\eps|$ (in linear-logarithmic coordinates) for all the instability mechanisms examined in this paper, both for anticyclonic ($\eps>0$) and  cyclonic ($\eps<0$) flows. The asymptotic estimates (solid lines) are compared with numerical results (symbols) for $s=0.1$. } \label{fig:comparison2analyt}
\end{center}
\end{figure}

Our results for all the types of instabilities are summarized by  
Figure \ref{fig:comparison2analyt}. This compares asymptotic and
numerically computed values of ${\Im} c$ as a function of $1/|\epsilon|$
for KW-KW, KW-IGW and IGW-IGW in anticyclonic flows,
and IGW-IGW instabilities in cyclonic flows.
The values of $\Im c$ displayed correspond to the maximum
over $m$ and $k$ for fixed $s=0.1$. 
For KW-KW instabilities, the asymptotic estimates are obtained
from \eqn{kwgr}. 
For KW-GW and IGW-IGW instabilities, we use  \eqn{kwigwro} and \eqn{igwigwro}, respectively.  These give $\Im c$ only up to a multiplicative constant which we fix by matching the asymptotic and numerical results for the smallest
values of $|\epsilon|$ shown in the figure. In the linear-logarithmic coordinates used, the numerical points line up with the predicted straight lines for larger $|\eps|$, thus confirming the validity of the asymptotic  analysis. Further support is provided  by the fact that the values of $k$ and $m$ for which $\Im c$ is maximised are close to the estimates  
\eqn{kwigwkm} and \eqn{igwigwkm}.
Evidently, the match
between the numerical and analytical results is quite good even
for $\eps$ moderately small. 
We see that the
instabilities become substantial for $\epsilon \approx 1$, especially
KW-KW instabilities. Observe that, as predicted by the analysis,
the decay of the growth rate in IGW-IGW instability as $\epsilon$
becomes small is the same for cyclonic and anticyclonic flows,
and yet the growth rates of cyclonic flow are smaller by a
factor of about 20. Thus the $O(1)$ prefactor in the asymptotics of $\Im c$ for IGW-IGW instabilities, ignored in \eqn{igwigwro}, turns out to be numerically very different for anticyclonic and cyclonic flows. The smallness of this prefactor in the cyclonic case means that the instability remains exceedingly weak even for $\eps \approx 1$, and likely irrelevant in many  physical situations. 

\section{Discussion} \label{sec:discussion}

This paper examines the linear stability of a horizontal Couette flow of a rapidly rotating, strongly stratified, inviscid fluid. The main conclusion is that the flow is unconditionally unstable: unbalanced instabilities, associated with linear resonances between Kelvin and inertia-gravity waves, occur for arbitrarily small Rossby numbers $\eps=\Lambda/f$. The growing perturbations have small horizontal and vertical scales, with typical wavenumbers or spatial-decay rates of the order of $\eps^{-1}$. 
Physically, it is easy to understand why asymptotically small scales are a key ingredient of the instabilities. The phase locking between different waves which underlies the instability mechanisms requires the wave phase speed to be comparable to the basic flow velocity, and this only occurs for small-scale waves.
The need for small vertical scales also explains why the instabilities examined in this paper have no direct counterparts in shallow-water flows; these are stable for small enough $|\eps|$ because of the inherent limitation in vertical structure imposed by the shallow-water approximation. 

Our conclusion that the rotating stratified Couette flow is always unstable is of course in sharp contrast with the one that may be drawn from balanced models. Regardless of their accuracy, which can be any power $\eps^n$, they predict the stability of flows without inflection points such as the Couette flow. 
There is no contradiction, however, since the growth rates found for the unbalanced instabilities are exponentially small in $|\eps|$. 
In practice, this exponential dependence means that the 
instabilities are exceedingly weak when $|\eps|$ is small, but can become important rather suddenly as $|\eps|$ increases towards $1$ and beyond. 
If the instabilities are to play a significant role in the breakdown of balance in geophysical flows, this will therefore be in a manner that is extremely sensitive to the Rossby number.

In the literature, most attention has been paid to anticyclonic flows, and in particular to the coupled Kelvin-wave instability occuring in these flows. Our results clarify that cyclonic flows are also unstable, through an instability mechanism involving coupled inertia-gravity waves. This mechanism is also active in anticyclonic flows where, along with the instability mode mixing Kelvin and inertia-gravity waves, it provides an alternative to the well studied instability due to Kelvin-wave resonance \citep[see][]{yavn-et-al,mole-et-al01}. The focus on anticyclonic flows and Kelvin-wave instabilities is justified in practice by the fact that the associated growth rate is much larger than those of the other instability mechanisms, exponentially larger in fact in the limit $\eps \to 0$. The instability of the cyclonic flows is especially weak. This weakness is not completely accounted for by the exponential dependence on $1/\eps$, since this is the same for both anticyclonic and cyclonic flows whilst the growth rates obtained numerically are very different.  We conclude, then, that the exponential dependence and the $O(1)$ prefactor conspire to make the instability of cyclonic flows extremely weak, even for moderate $|\eps|$. 

The WKB approach used in this paper could be extended to examine the instability in more general rotating stratified shear flows. Obvious applications are the stratified Taylor--Couette flow \citep{yavn-et-al,mole-et-al01}, which differs from the problem studied here by the presence of curvature terms, and the stability of accretion discs \citep{rudi-et-al,dubr-et-al}. 
Additional physical effects that it would be of interest to study include
different boundary conditions (in particular the case of infinite domains for which no Kelvin waves exist), viscous and thermal damping, and non-zero potential-vorticity gradients, leading to the existence of critical levels for neutral modes \citep[cf.][]{balm99}.

JV was funded by a NERC Advanced Research Fellowship.

\appendix
\section{Conservation laws} \label{app:a}
Let
\[
M = (u \partial_z \rho - w \partial_x \rho)/N^2.
\]
Denoting integration over the periodic domain in $x$ and $z$ by
\[
\av{\cdot} = \int \! \! \int \cdot \, \, \d x \d z,
\]
we compute
\begin{eqnarray} \lab{flux}
N^2 \partial_t \av{M} &=& \av{\partial_t u \partial_z \rho - \partial_z u
  \partial_t \rho - \partial_t w \partial_x \rho + \partial_x w
  \partial_t \rho} \nonumber \\
&=& \av{ (f - \Lambda) v \partial_z
  \rho - \partial_x p \partial_z \rho - N^2 \partial_z u w +
  \partial_z p \partial_x \rho} \\
&=& - N^2 \partial_y \av{uv}, \nonumber
\end{eqnarray}
where we have used integration by parts and periodicity extensively,
and, for the last line, $q=0$ and the incompressibility equation.
The conservation for the quadratic wave momentum (or pseudomomentum)
\[
\mathcal{M} = \int \! \! \int \! \! \int ( u
\partial_z \rho - w \partial_x \rho ) \,  \d x \d y \d z /N^2
\]
follows by integration in $y$, using the boundary conditions $v=0$.

The perturbation energy $\mathcal{E}'$, with density $|\mathbf{u}|^2/2 +
\rho^2/(2N^2)$, is not conserved but satisfies
\[
\dt{\mathcal{E}'}{t} =  - \int \! \! \int \! \! \int \Lambda u v \, \d
x \d y \d z.
\]
Integrating by parts the right-hand side and using \eqn{flux} gives a
conservation law for the wave energy (or pseudoenergy)
\[
\mathcal{E} = \int \! \! \int \! \! \int \left(\frac{|\mathbf{u}|^2}{2} +
\frac{\rho^2}{2 N^2} +
\Lambda y \frac{u \rho_z  -w \rho_x }{N^2} \right) \, \d x \d y \d z.
\]
Note that the conservation of both $\mathcal{M}$ and $\mathcal{E}$ can also
be derived from the exact conservation laws for momentum, energy
and potential vorticity for the full system, that is, basic flow plus
perturbation.

\section{Equation for $\hat{u}$} \label{app:u}

In \S2, the eigenvalue problem satisfied by normal-mode solutions is
formulated as the second-order differential equation \eqn{k1} for
$\hat p$ and its associated boundary condition \eqn{bc}
\citep[cf.][]{kush-et-al}. An alternative formulation, employed by
\citet{yavn-et-al}, uses $\hat u$ instead of $\hat p$ as the dependent
variables. It has the advantage that the removable singularities that
appear in \eqn{k1} are absent. For completeness, we record this
alternative formulation as
\beq \lab{u}
\eps^2 \left( \frac{1-s^2 \hom^2}{K} \hat{u}' \right)' - \left(
  \frac{k^2(1-s^2 \hom^2) + m^2 (1-\eps-\hom^2)}{K} + \frac{2\eps(1-\eps) s^2
    k^2 m^2 \hom^2}{K^2} \right) \hat u = 0,
\eeq
where
\[
K = (1-s^2 \hom^2) k^2 + (1-\eps)^2 m^2.
\]
The associated boundary conditions are \beq \lab{bcu} \eps
\hat{u}' + \frac{(1-\eps) m^2 \hom}{k(1-s^2 \hom^2)}\hat u = 0
\quad \textrm{at} \ \ y=\pm 1. \eeq This is the formulation used
for the numerical computation of the normal modes.

\section{Kelvin-wave dispersion relation} \label{app:kwinst}

In this Appendix, we derive the dispersion relation for KWs accurate
to $O(\eps)$, as is necessary to obtain the leading-order asymptotics
of the KW-instability growth rate.

The dispersion relation for KW$_\pm$ valid to all orders in $\eps$ are
given in \eqn{dispex}. It is solved at leading order in \S\ref{sec:kelvin} to give
\eqn{c0}. At the next order, we find the two equations
\beq \lab{disp1}
\pm \sigma c_1 \lambda(\pm 1) g_{\pm}( \pm 1) + \hc_0(\pm 1) g_{\pm}'(\pm1) = 0,
\eeq
which allow the determination of the $O(\eps)$ contribtion to the
frequency $\omega_1$. Note that the
contributions of the $O(\eps)$ terms neglected in \eqn{p-}--\eqn{p}
cancel in these two equations when \eqn{om0} is taken into account.
Equation \eqn{g'/g} can be used to express the derivatives of
$g_{\pm}$; the following results are therefore useful:
\begin{eqnarray}
\lambda(\pm 1)&=& r, \lab{lampm1} \\
-\frac{\lambda'(\pm 1)}{2\lambda} &=&
  \frac{\pm \sigma (1-s^2)k^2 r}{m^2}, \nonumber \\
\frac{k \hom_0{(\pm1)}}{1-\hom_0^2(\pm1)} &=&
\frac{\mp \sigma k^2 r}{r^2-k^2}, \nonumber  \\
\frac{\mp \sigma h(\pm1)}{2 \lambda(\pm1)}&=& \mp \sigma r
   \left(\frac{1}{2}-\frac{k^2}{r^2-k^2}\right) +
c_1 \frac{(1-s^2)k^2 r^2}{m^2}. \nonumber
\end{eqnarray}
Using these and \eqn{g'/g}, \eqn{disp1} gives the first-order
correction to the frequencies \eqn{om0},
\beq \lab{c1}
c_1 = \frac{\mp \sigma}{2 r}.
\eeq

\bibliographystyle{agsm}
\bibliography{mybib}

\end{document}